\documentstyle[12pt]{article}
\begin{document}

\centerline{\bf Extraction of the D$_{13}$(1520) photon-decay couplings}
\centerline{\bf from pion- and eta-photoproduction data} 
\vskip .3cm
\centerline{Ron Workman, Richard A. Arndt, and Igor I. Strakovsky} 
\centerline{Center for Nuclear Studies, Department of Physics}
\centerline{The George Washington University, Washington, DC 20052}

\begin{abstract}

We compare results for the D$_{13}$(1520) photon-decay amplitudes determined
in analyses of eta- and pion-photoproduction data. The ratio of helicity
amplitudes (A$_{3/2}$/A$_{1/2}$), determined from eta-photoproduction data,
is quite different from that determined in previous analyses of 
pion-photoproduction data. We consider how strongly the existing 
pion-photoproduction data constrain 
both this ratio and the individual photon-decay amplitudes.

\end{abstract}

\vskip .5cm

Recent precise measurements of eta-photoproduction observables have 
spawned a number of analyses, focused mainly on the properties of the
S$_{11}$(1535) resonance\cite{krusche}. 
These studies have found values for the 
photo-decay amplitude, $A_{1/2}^p$, which are significantly larger than
those found in previous analyses of pion-photoproduction 
data\cite{amplitude_s11}. As the 
S$_{11}$(1535) resonance is masked by a strong $\eta N$ threshold
cusp in pion photoproduction, eta-photoproduction holds the promise of
a less model-dependent analysis. Attempts to fit pion- and eta-production
data in coupled-channel approachs\cite{dytman,feuster} have generally
found values between those extracted from single-channel fits.   

Two studies\cite{mukhopadhyay,mainz}
have gone beyond the S$_{11}$(1535) and have considered the
sensitivity of eta-photoproduction data to the nearby D$_{13}$(1520) 
resonance. In both of these analyses, values for the
{\it ratio} of photo-decay amplitudes, A$_{3/2}$/A$_{1/2}$, were found
to be consistently far smaller than those inferred from pion-photoproduction
analyses\cite{amplitude_d13}. 
This discrepancy is certainly unexpected\cite{expect}, 
as the D$_{13}$ state appears to have
a clean Breit-Wigner-like signal in the
associated multipoles E$_{2-}^{1/2}$ and M$_{2-}^{1/2}$ extracted
from pion-photoproduction data. 
This ratio, as determined from eta-photoproduction data, has the value 
$-2.5\pm 0.2 \pm 0.4$\cite{mukhopadhyay} or $-2.1\pm 0.2$\cite{mainz},
as compared to the PDG estimate\cite{PDG} of $-6.9\pm 2.1$. 
Eta-photoproduction has the advantage of isospin selectivity but, in the
case of the D$_{13}$(1520), one must deal with a very small coupling to
the $\eta N$ channel. 

As this difference amounts to a shift by several standard deviations in
a supposedly well-determined quantity, we have considered whether the 
eta-photoproduction result can be accommodated, even qualitatively, by
the existing pion-photoproduction database. Given that we are investigating
a very large effect, and the background contribution to the 
E$_{2-}^{1/2}$ and M$_{2-}^{1/2}$ multipoles appears to be small near
the resonance energy, this study was carried out assuming resonance
dominance in both the eta- and pion-photoproduction multipoles\cite{comment}. 
Clearly this implies our results will only be qualitative. 
However, as we will see, even qualitative results can be revealing. 

We first note that the ratio of modified multipole amplitudes,
corresponding to the ratio A$_{3/2}$/A$_{1/2}$ is given by
\begin{equation}
{ A_{3/2} \over { A_{1/2} }} = \sqrt{3} \left( { { \bar E_{2-} + 
  \bar M_{2-} } \over { \bar E_{2-} - 3\bar M_{2-} } } \right) 
\end{equation}
with conversion factors as given in, for example, Ref.\cite{vpi90}. Since
we will be dealing with ratios, the conversion factor is not relevant and
we will drop the barred notation. Assuming resonance dominance, a ratio
of $-2.5$ for A$_{3/2}$/A$_{1/2}$ can be converted to a ratio of about
$1.4$ for E$_{2-}^{1/2}$/M$_{2-}^{1/2}$. This can be compared to the
result of a representative analysis of pion-photoproduction 
data\cite{sm99}, wherein the ratio of multipoles (imaginary
parts) is found to be about 2.1 at the resonance energy\cite{phase}.
 
In order to gauge the sensitivity of pion-photoproduction data to
this ratio,
we started with a single-energy analysis centered at a lab
photon energy of 760 MeV, corresponding to a value of $\sqrt{s}$ near 
the D$_{13}$(1520) resonance position. 
We then considered the effect of changes in the fitted multipoles. Some
qualitative results were immediately noticed.
If one D$_{13}$ multipole was fixed and
the other (E$_{2-}^{1/2}$ or M$_{2-}^{1/2}$) was shifted to achieve a
ratio of 1.4, the cross sections for both $\pi^0 p$ and $\pi^+ n$
production were missed by large margins. However, if E$_{2-}^{1/2}$ was
reduced and M$_{2-}^{1/2}$ was increased in magnitude, a qualitative
description of the cross sections could be retained. A good fit
to the existing polarization data was also preserved. It soon became
apparent that a {\it small} increase in M$_{2-}^{1/2}$ and a {\it moderate}
decrease in E$_{2-}^{1/2}$ was preferred in this exercise. From 
Eq.~(1) this implies a small decrease in A$_{3/2}$ and a larger 
increase in the magnitude of A$_{1/2}$; results following the trend 
suggested in Refs.\cite{mukhopadhyay,mainz}.

In Fig.~1 we show the result of increasing M$_{2-}^{1/2}$ by 15\% and
fixing the E$_{2-}^{1/2}$/M$_{2-}^{1/2}$ ratio at 1.4. 
For comparison purposes, we also show the result of a shift in
M$_{2-}^{1/2}$ alone, leading to the required ratio. 
The backward-angle cross sections are particularly sensitive to these
changes, as the D$_{13}$ multipoles enter in the combination
(E$_{2-}^{1/2}$ $-$ 3M$_{2-}^{1/2}$). The larger M$_{2-}^{1/2}$ and
smaller E$_{2-}^{1/2}$ both reduce the cross section at back angles.
Polarization measurements, in the current data base, are not 
sufficiently precise to pin down E$_{2-}^{1/2}$ and M$_{2-}^{1/2}$.
The relative insensitivity of recoil polarization is illustrated 
in Fig.~2. A somewhat greater sensitivity is seen in the 
beam-polarization observable ($\Sigma$). It should be
emphasized that this is {\it not} a fit to the pion-photoproduction data.
As mentioned above, a fit would result in very different values for the
multipoles. Here we are simply showing how the conclusions of 
Refs.\cite{mukhopadhyay,mainz} would effect the existing fit to 
pion-photoproduction data, near the D$_{13}$ resonance position.

In summary, properties of the D$_{13}$ multipoles, as determined from
fits to eta-photoproduction data, are not entiredly excluded by the existing
pion-photoproduction data. It is possible to obtain a qualitative description
(but not a $\chi^2$ fit)
of the pion-production data, at the resonance position, consistent with an
A$_{3/2}$/A$_{1/2}$ ratio near $-2.5$. If this ratio were correct, and effects
from the background were not a problem, the next step would be to determine
which data, in the pion data base, were incompatible with this result. At 
present this would be difficult, as the data base is rather sparse, with
few sets covering a wide angular range. We suggest a similar study should
be interesting if performed on the eta-photoproduction data base. In that
test one would assume the D$_{13}$ ratio, as extracted from pion-production
data, and consider what changes in the other multipoles would be required
for a qualitative fit.

A second implicit assumption in this
study should also be mentioned. We have considered the effect of changes in
the D$_{13}$ multipoles {\it assuming} the remaining multipoles to be 
correct. Given the above mentioned discrepancy between eta- and 
pion-photoproduction results for the S$_{11}$(1535), this assumption could
be questionable for the S-wave multipoles. As an exercise, we fixed the
E$_{2-}^{1/2}$/M$_{2-}^{1/2}$ ratio at the resonance point, and fit the
full database to 1.2 GeV. In this fit, as might be expected, the
E$_{0+}^{1/2}$ multipole showed the largest shift due to the constraint. 
More definitive tests will be possible when precise measurements of the
cross section and polarization observables cover this region. Precise
measurements at backward angles will be particularly useful.

The authors thank B. Krusche and J. Ahrens for providing preliminary
data from Mainz. R.W. thanks C.~Bennhold for useful comments on the
eta-photoproduction analyses.
This work was supported in part by a U.S. Department of Energy Grant No.
DE-FG02-99ER41110. R.W. and I.S. gratefully acknowledge a contract from
Jefferson Lab under which this work was done. The Thomas Jefferson 
National Accerator Facility (Jefferson Lab) is operated by the 
Southeastern Universities Research Association (SURA) under DOE 
contract DE-AC05-84ER40150.

\eject

\centerline{FIGURE CAPTIONS}
\parindent=0pt

\vskip 2cm

Figure 1. Differential cross section for $\gamma p\to p\pi^0$ at 762 MeV.
Data from Ref.\cite{mainz_pion}. Solid curve corresponds to the unmodified
single-energy solution; dashed curve corresponds to a 15\% increase in
the imaginary part of M$_{2-}^{1/2}$ and the proposed A$_{3/2}$/A$_{1/2}$
ratio (-2.5); dotted curve corresponds to the proposed ratio assuming 
the largest multipole (E$_{2-}^{1/2}$) is correct.\hfil\break

Figure 2. Recoil polarization for $\gamma p\to p\pi^0$ at 762 MeV. Data,
between 760 and 765 MeV, from Ref.\cite{recoil}. Curves as given in Fig.~1.


\begin{thebibliography}{99}
\bibitem{amplitude_s11}
In GeV$^{-1/2}$ units, typical pion-production results for the 
S$_{11}$(1535) A$^{p}_{1/2}$ are 0.060$\pm 0.015$ (VPI) and
0.053$\pm 0.15$ (Glasgow). Results from Ref.\cite{krusche} 
are in the range 0.120$\pm 0.011 \pm 0.015$.
\bibitem{amplitude_d13}
Typical values, in GeV$^{-1/2}$ units, for the D$_{13}$(1520) 
A$_{3/2}^p$ and A$_{1/2}^p$ are respectively 0.167$\pm 0.005$ and 
-0.020$\pm 0.007$ (VPI), 0.156$\pm 0.022$ and \hbox{-0.028$\pm 0.014$}
(Glasgow).
\bibitem{expect}
Phenomenological values of the helicity 1/2 and 3/2 transition
amplitudes have remained relatively stable in fits 
to pion-photoproduction data.
Quark models, however, have predicted ratios ranging between the pion-
and eta-photoproduction results. See Ref.\cite{mukhopadhyay} for a 
discussion of the various quark model predictions.
\bibitem{comment} 
A small background is expected in eta-photoproduction based on the
small $\eta$NN coupling constant. The justification for this approximation
in pion-photoproduction is more model-dependent.
Comparing the ratio of helicity 1/2 and 3/2 transition amplitudes, 
calculated in our full analysis, versus an estimate based on 
E$_{2-}^{1/2}$ and M$_{2-}^{1/2}$ taken at the resonance position 
(no background separation), we find agreement at the 10-20\% level.
If an improper background-resonance separation should be the problem in
this case, the uncertainties associated with all other states would be
questionable as well.
\bibitem{phase}
This study is helped by the observation that the E$_{2-}^{1/2}$ and
M$_{2-}^{1/2}$ multipoles, as extracted from pion-photoproduction data,
have very nearly equal phases which pass
through 90$^\circ$ very near the resonance energy.
\end{thebibliography}
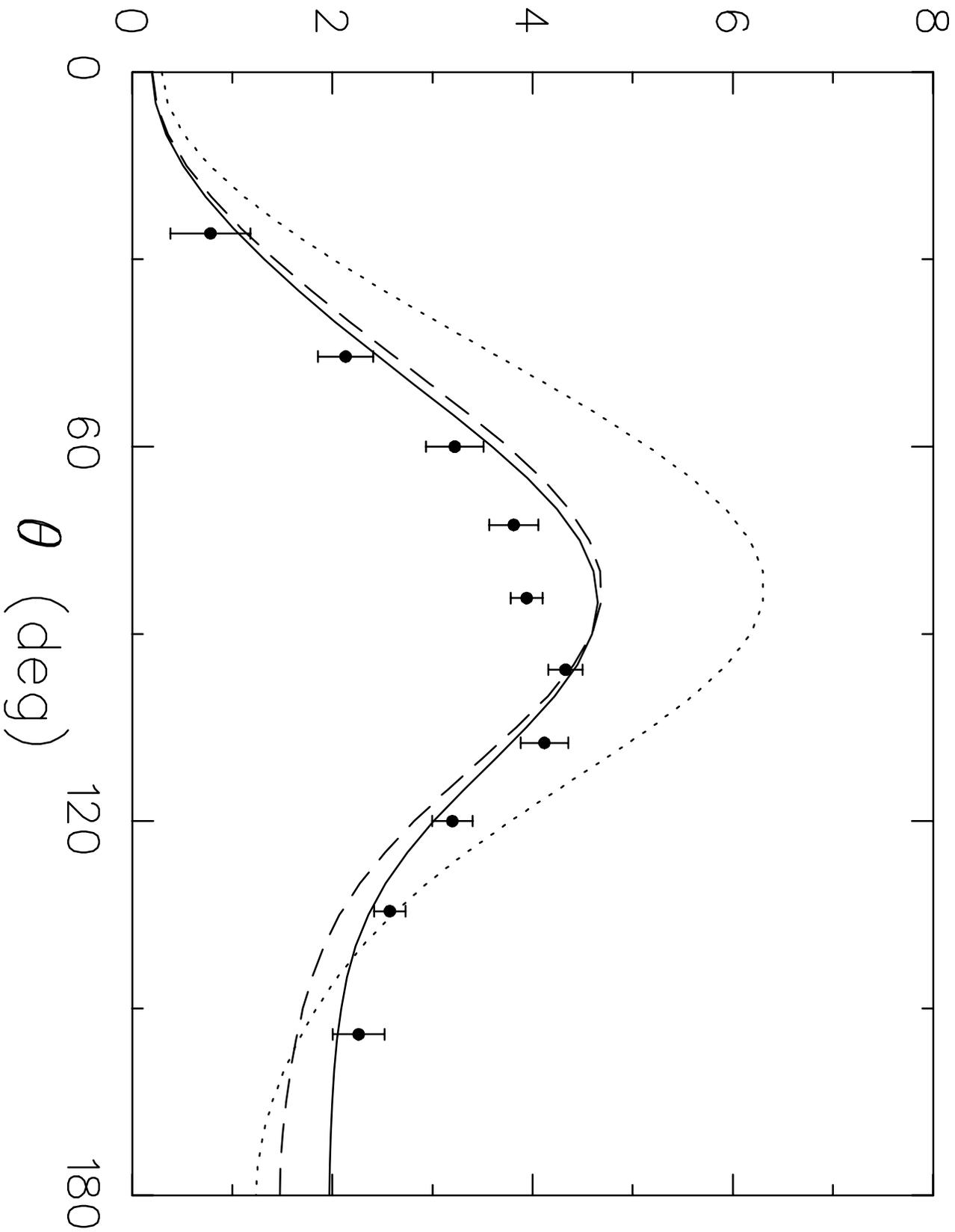
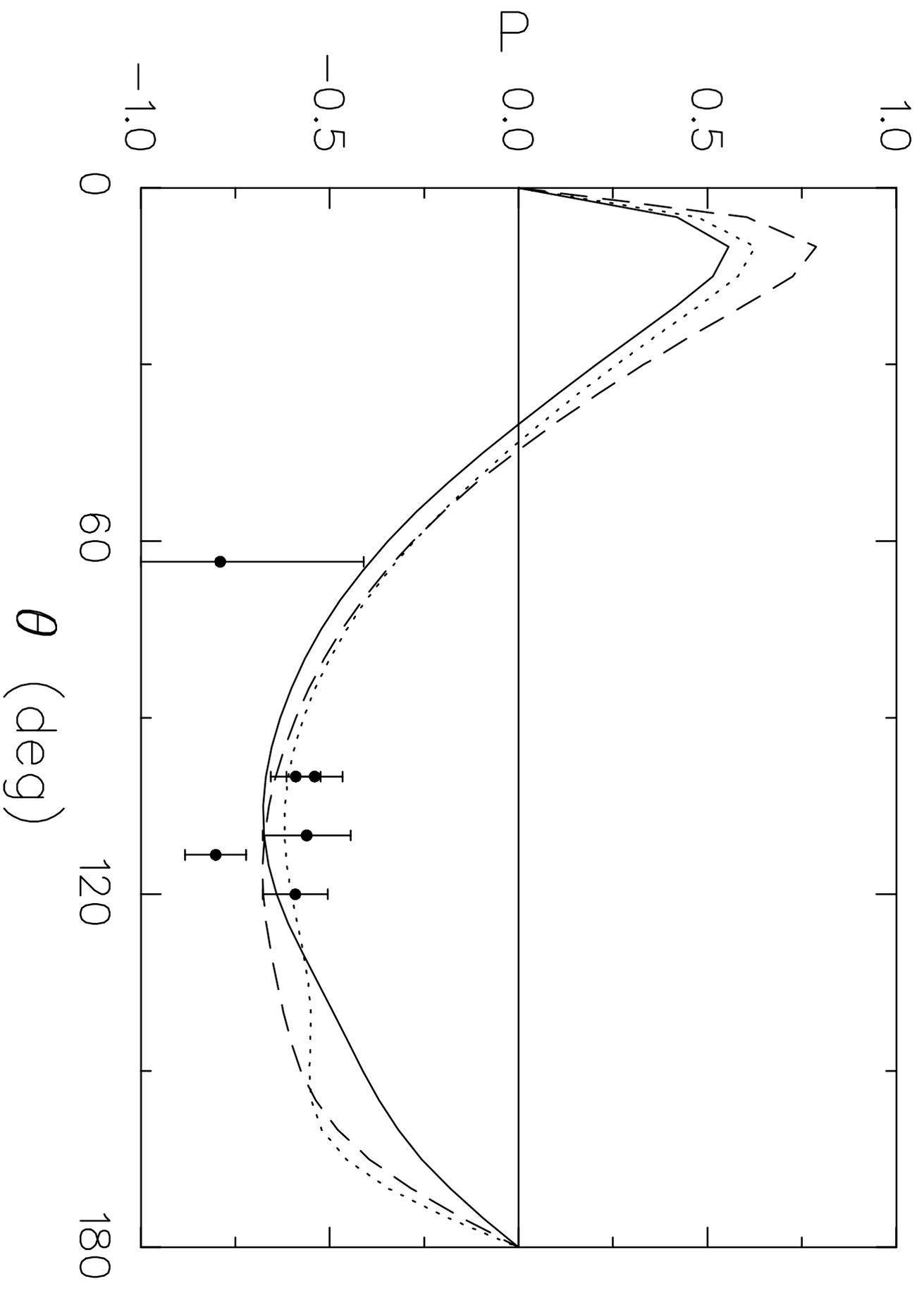
\end{document}